\documentclass[conference]{IEEEtran}
\IEEEoverridecommandlockouts
\usepackage{cite}
\usepackage{amsmath,amssymb,amsfonts}
\usepackage{graphicx}
\usepackage{textcomp}
\usepackage{subfigure}
  
\usepackage{tabularx}
\usepackage{hyperref}
\usepackage[ruled,linesnumbered]{algorithm2e}
\usepackage{algpseudocode}
\usepackage{booktabs}
\usepackage[inline]{enumitem}
\usepackage{mathtools}
\usepackage{verbatim}
\usepackage{tikz}

\newcommand*\circled[1]{\tikz[baseline=(char.base)]{
            \node[shape=circle,draw,inner sep=2pt] (char) {#1};}}

\def\BibTeX{{\rm B\kern-.05em{\sc i\kern-.025em b}\kern-.08em
    T\kern-.1667em\lower.7ex\hbox{E}\kern-.125emX}}

\begin{document}

\title{EdgeChain: Blockchain-based Multi-vendor\\Mobile Edge Application Placement}

\author{
	\IEEEauthorblockN{He Zhu}
	\IEEEauthorblockA{\textit{Systems and Computer Engineering} \\
	\textit{Carleton University}\\
	Ottawa, ON, Canada \\
	hzhu@sce.carleton.ca}
\and
    \IEEEauthorblockN{Changcheng Huang}
	\IEEEauthorblockA{\textit{Systems and Computer Engineering} \\
	\textit{Carleton University}\\
	Ottawa, ON, Canada \\
	huang@sce.carleton.ca}
\and
	\IEEEauthorblockN{Jiayu Zhou}
	\IEEEauthorblockA{\textit{Computer Science and Engineering} \\
	\textit{Michigan State University}\\
	East Lansing, MI, USA \\
	jiayuz@msu.edu}
}

\maketitle

\begin{abstract}
The state-of-the-art mobile edge applications are generating intense traffic and posing rigorous latency requirements to service providers. While resource sharing across multiple service providers can be a way to maximize the utilization of limited resources at the network edge, it requires a centralized repository maintained by all parties for service providers to share status. Moreover, service providers have to trust each other for resource allocation fairness, which is difficult because of potential conflicts of interest. We propose \textsf{EdgeChain}, a blockchain-based architecture to make mobile edge application placement decisions for multiple service providers. We first formulate a stochastic programming problem minimizing the placement cost for mobile edge application placement scenarios. Based on our model, we present a heuristic mobile edge application placement algorithm. As a decentralized public ledger, the blockchain then takes the logic of our algorithm as the smart contract, with the consideration of resources from all mobile edge hosts participating in the system. The algorithm is agreed by all parties and the results will only be accepted by majority of the mining nodes on the blockchain. When a placement decision is made, an edge host meeting the consumer's latency and budget requirements will be selected at the lowest cost. All placement transactions are stored on the blockchain and are traceable by every mobile edge service provider and application vendor who consumes resources at the mobile edge.
\end{abstract}


\section{Introduction}
\label{sec:intro}
The rapid advance of mobile edge computing (MEC) has been the last mile of enabling a shared, low-latency computational environment for multi-vendor mobile edge applications. MEC performs computing offloading, data storage, caching and processing, request distribution and service delivery from the mobile edge to end users \cite{shi2016edgevision}. Applications with low latency tolerance, such as augmented reality (AR), video streaming, and online gaming, can deploy their services on the edge hosts at a cost, to achieve lower latency and better user experience \cite{ec-etsi-hu15}.

As the market gets mature, there will be multiple 5G service providers (SPs) provisioning MEC services to cover the same area: bigger wholesale players will invest in infrastructure to actually build mobile edge base stations, while there will also be mobile virtual network operators (MVNOs) renting resources from the former. These SPs can collaborate with each other in several ways for better utilization of the resources at the edge: virtual SPs have to place mobile edge (ME) applications on one of the rented edge hosts, preferably with lower cost, regardless of SPs. On the other hand, MEC base stations from different SPs can share resources with each other to process bursting requests.

\begin{figure*}[t]
\centering
\includegraphics[width=.9\textwidth]{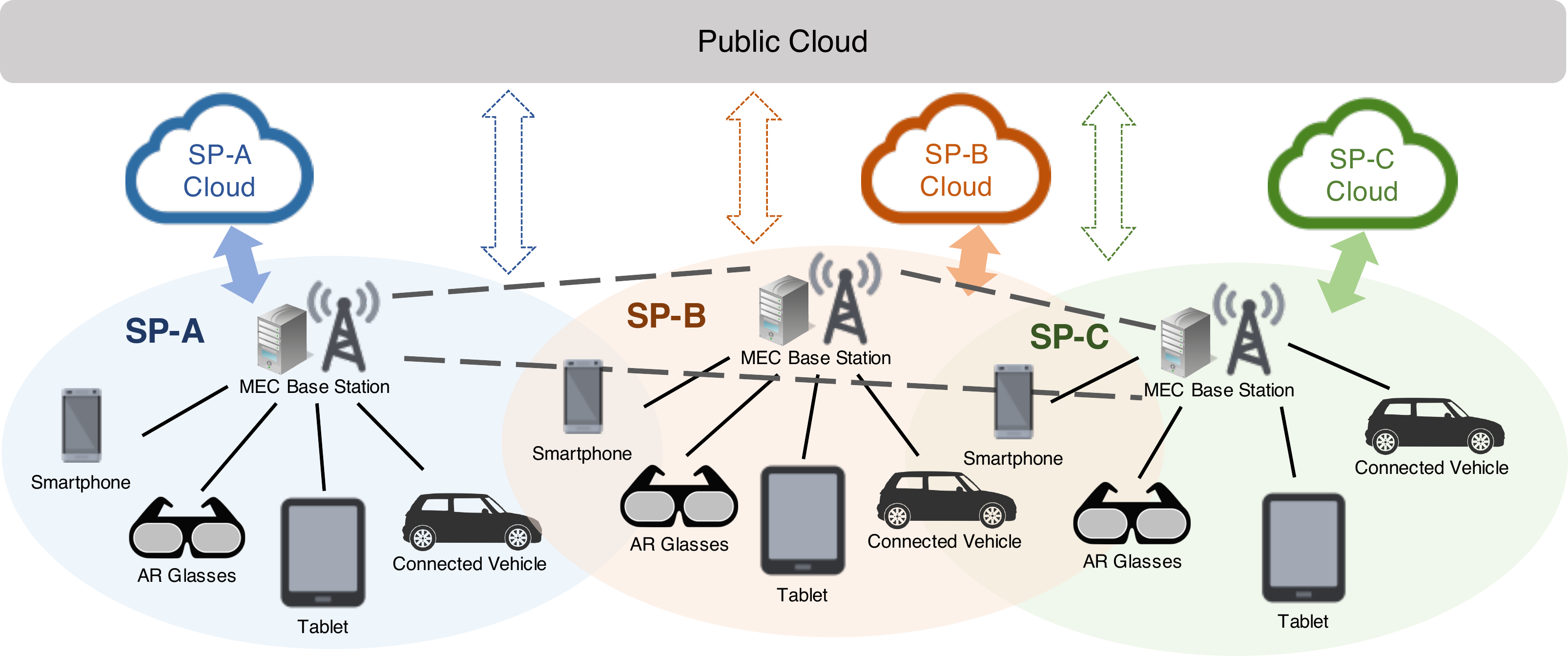}
\caption{\textbf{A MEC scenario in a certain service area. There are 3 ME base stations from 3 different SPs: SP-A, SP-B, and SP-C. They serve there own users within the service area. For resource sharing and optimization purpose, the base stations are also connected to each other.}}
\label{fig:intro}
\end{figure*}

For encouraging SPs to enroll their eligible MEC base stations and hosts in resource sharing, it is common to give incentives to SPs for contributing their resources of the hosts for hosting edge applications. Following the changing demand of end users, certain types of edge applications need to be deployed on, migrated to, or removed from an edge host, in order to meet the service requirement. By deploying the edge applications at the right places, the edge application provider will save costs, while providing high-quality service with low latency to the end users. Meanwhile, the edge host will collect incentives for its resources effectively used.

Clearly, the edge computing framework needs a placement service to dynamically check the user needs and the available edge hosts, and determine the placement or removal of edge applications. In datacenters, virtual machine (VM) placement has been well investigated, mainly with the focus of more efficient resource utilization and lower operational expense (OPEX). However, the collaboration of multiple SPs and mobile edge applications vendors are posing new challenges for ME application placement from the following aspects:

\begin{itemize}
	\item A placement model has to make transparent and consistent selections of the best host for each request for edge computing resources. Moreover, the model has to take into consideration that a mobile edge application may require multiple services chained together at the edge.
	\item A trusted party is required to determine the best place for application deployment. When an edge application is deployed on a mobile edge host, the application vendor needs to pay for the usage of the host. The placement algorithm has to avoid affiliation to either SP to ensure a neutral decision is made strictly according to the resource and the cost. It may create conflicts of interest to put any SP involved into the position of making placement decisions: placing mobile edge applications onto the SP's own hosts would bring revenue for renting their resources.
	\item The application placement service needs to be steadily available. Both the mobile edge hosting service providers and the mobile edge application providers can constantly change. The placement service provider must remain in service regardless of the joining or quitting of vendors.
\end{itemize}

The challenges above urge a comprehensive solution uniting all SPs and their edge hosts without bias. In this paper, we present an architecture combined with its algorithm, namely \textsf{EdgeChain}, to create a decentralized placement service for mobile edge application that does not require trust to any party, i.e., trustless placement service. Compared with current placement solutions, \textsf{EdgeChain} has the following contributions:
\begin{itemize}
	\item A cost model is presented as a stochastic programming problem, factoring in the pricing of edge hosts, latency, and service chaining.
	\item We develop a heuristic placement algorithm based on the proposed cost model with the consideration of efficiency for running by the blockchain.
  \item We introduce blockchain technologies to the MEC resource orchestration framework with two considerations: the first is to store the global resource availability, allocation, and consumption information that helps our algorithm make optimized decisions based on the global resource information. The second consideration is to have a decentralized public ledge for ensuring the neutrality of the placement decisions.
	\item The \textsf{EdgeChain} framework is presented to run our algorithm for making placement decisions. In our design, SPs and mobile edge application vendors participate in the maintenance of the blockchain. An \textsf{EdgeChain} client is embedded in the network function virtualization (NFV) framework to determine the placement based on the existing information on the blockchain. To our best knowledge, this is the first work that leverages blockchain to coordinate SPs for MEC application placement.
  \item Simulation results of our placement algorithm show its effectiveness in mobile edge host resource sharing among SPs. We also implement the \textsf{EdgeChain} by leveraging VeChain \cite{vechain}, an enterprise-level blockchain-as-a-service framework derived from Ethereum \cite{wood2014ethereum}. 
\end{itemize}

We divide the contents into the following sections. The related work is illustrated in Section \ref{sec:relatedwork}. Section \ref{sec:problem} formulates the problem. Section \ref{sec:algorithm} proposes the heuristic \textsf{EdgeChain} placement algorithm based on the problem formulation. Then the simulation results are shown in Section \ref{sec:numerical}. Section \ref{sec:conclusion} concludes the paper.

\section{Related Work}
\label{sec:relatedwork}
The research directions in network service chaining (NSC) were discussed in \cite{john2013sfc-direction}. For security considerations, the authors highlighted the difficulty of bringing short-lived network services to targeted users in a single subscriber network by using the current security schemes. The potential security problems in SFC were stated in RFC7498 \cite{quinn2015problem}, including service overlay security, trusted classification policy, and secure SFC encapsulation. We investigated a placement problem in MEC with the consideration of application availability in \cite{zhu2017availability}.

Xiong et al. proposed a pricing strategy for offloading the blockchain's resource-consuming proof-of-work tasks to edge computing nodes \cite{xiong2017edge}. A two-stage Stackelberg game model was presented with both the edge computing service provider and the miners involved. A hierarchical distributed control system was built using Hyperledger Fabric blockchain \cite{stanciu2017control}. The hosting locations of cloud and fog of blockchain were compared in \cite{samaniego2017baas} for IoT networks with the conclusion that fog nodes were better as network latency was the dominant factor.

Nakamoto introduced the concept of blockchain and implemented Bitcoin \cite{nakamoto2008bitcoin}, a decentralized cryptocurrency that first resolved the double spending problem. Blockchains are based on Merkle trees \cite{Bayer1993} to efficiently allow multiple documents to be saved together in a block. As a decentralized public ledger, blockchains can serve beyond cryptocurrencies. Ethereum \cite{wood2014ethereum} used blockchain to store smart contracts that support building virtually any decentralized application.

\begin{table}[tb]
  \centering
  \caption{Parties involved in a MEC placement scenario}
  \begin{tabularx}{.48\textwidth}{p{.07\textwidth}X}
  \toprule %
  \textbf{Party}             &  \textbf{Description}  \\
  \toprule %
  \textit{Users}             &  Subscribers of applications and services over 5G networks with MEC enabled. \\
  \textit{MECSPs}            &  MEC service providers, who deliver MEC hosting services that can run \textit{MEApps} at the network edge, close to end users. Examples include telecommunication companies like Rogers and Telus in Canada. \\
  \textit{MEAVs}             &  Mobile edge application vendors, who provide \textit{MEApps} and services to end users. For instance, a company selling AR services. \\
  \textit{MEApps}            &  MEApps stand for mobile edge applications provided by MEAVs. \\
  \textit{MEHosts}           &  Servers that belong to different MECSPs to provide hosting service of \textit{MEApps}. \\
  \textit{HostLinks}        &  Network links between hosts, regardless of which MECSP they belong to. \\
  \textit{AppLinks} &  When \textit{MEApps} are chained together, virtual links will be established for data transmissions traveling through the chain. \\
  \bottomrule
  \end{tabularx}
  \label{table:parties}
\end{table}

\begin{table}[tb]
  \centering
  \caption{Notations Used in Problem Formulation}
  \begin{tabularx}{.48\textwidth}{lX}
  \toprule %
  \textbf{Notation}                 &  \textbf{Description}  \\
  \toprule %
  $s, \mathbb{V}_s, \mathbb{L}_s, v, l$ &  $s$ is a service chain. $\mathbb{V}_s$ is the set of all \textit{MEApps} in $s$. $\mathbb{L}_s$ is the set of all \textit{AppLinks} in $s$. A \textit{MEApp} in $s$ is denoted by $v \in \mathbb{V}_s$, and an \textit{AppLink} between two \textit{MEApps} in $s$ is denoted by $l \in \mathbb{L}_s$. \\
  $\mathbb{H}, \mathbb{E}, h, e, \mathbb{V}_h$                 &  $\mathbb{H}$ is the set of all \textit{MEHosts}. $\mathbb{E}$ is the set of all \textit{HostLinks}. A \textit{MEHost} is denoted by $h \in \mathbb{H}$, and a link between two \textit{MEHosts} is denoted by $e \in \mathbb{E}$. $\mathbb{V}_h$ is the set of all \textit{MEApps} placed on $h$. \\
  $u, m, c_s$                    &   $u$ is an end user. $m$ is a MECSP. $c_s$ is the cost of deploying $s$. \\
  $c_v, c_{v h_{i}, v' h_{j}}$      &  $c_v$ is the cost of deploying $v$. $c_{v h_{i}, v' h_{j}}$ is the cost of the \textit{AppLink} between $v$ on $h_i$ and $v'$ on $h_j$. \\
  $n_s, P_m, h_m$                   &  $n_s$ is the total number of users requesting $s$. $P_m$ is a random variable denoting the percentage of the users of MECSP $m$. $h_m$ is an edge host of $m$. \\
  $\gamma_m, \delta_m$              &  $\gamma_m$ is the unit price of serving $m$'s own subscribers. $\delta_m$ is the extra charge for $m$ serving users of other MECSPs. \\
  $C_v, M_v$                        & CPU and memory requirement of the \textit{MEApp} $v$. \\
  $B(e_{ij}), \zeta_{e_{ij}}$       & $B(e_{ij})$ is the total bandwidth capacity of \textit{HostLink} $e_{ij}$. $\zeta_{e_{ij}}$ is the unit price of the bandwidth of $e_{ij}$.\\
  $B_V(e_{ij})$                     & $B_V(e_{ij})$ is the total bandwidth used by \textit{MEApps} deployed on $h_i$ and $h_j$. \\
  $B(v_{h_i}, v_{h_j})$             & Bandwidth used between \textit{MEHosts} $h_i$ and $h_j$. \\
  $C_h, M_h$                        & CPU and memory capacity of the \textit{MEHost} $h$. \\
  $t_{e_{ij}}, t_s, T_s$            & $t_{e_{ij}}$ is the latency incurred on \textit{HostLink} $e_{ij}$. $t_s$ is the latency of the service chain $s$. $T_s$ is the max latency allowed by $s$. \\
  \bottomrule
  \end{tabularx}
  \label{table:notations}
\end{table}

\section{Problem Formulation}
\label{sec:problem}
We first list all parties involved in a MEC placement scenario in Table \ref{table:parties}. The problem is formulated from a \textit{MEAV}'s point of view: \textit{MEApps} are direct consumers of the computing resources in the MEC environment, because a \textit{MEAV} needs to pay \textit{MECSPs} for hosting its applications in order to serve their \textit{users} and meet the latency requirement. Each \textit{MEApp} is equivalent to a virtual machine (VM) deployed on a \textit{MEHost}. \textit{MEApps} provided by different \textit{MEAV} can be combined as a service chain to provide comprehensive services. A service chain may span multiple \textit{MEAVs}. In this case, revenues generated by the service chain can be distributed according to the usage of each \textit{MEApp} on the service chain. For instance, a full-fledged AR service can load real-time navigation information from an online map application, while it can also load promotions of a shopping mall nearby from the mall's application. The navigation data is collected by the online map application, and the shopping mall application gets paid if the user "clicks" the links of the promotions.

The notations used in formulating the problem is shown in Table \ref{table:notations}. Define a chained service $s$ as a forwarding graph \cite{brown2015service} $G_s = (\mathbb{V}_s, \mathbb{L}_s)$, where $\mathbb{V}_s$ is the set of all \textit{MEApps} contributing to the service, and $\mathbb{L}_s$ is the set of all \textit{AppLinks} connecting applications together. A \textit{MEApp} is denoted by $v \in \mathbb{V}_s$, and an \textit{AppLink} between two \textit{MEApps} is denoted by $l \in \mathbb{L}_s$.

The chained service is deployed on a graph of connected \textit{MEHosts} $G_h = (\mathbb{H}, \mathbb{E})$, where $\mathbb{H}$ is the set of all \textit{MEHosts} owned by various \textit{MECSPs} and $\mathbb{E}$ is the set of all \textit{HostLinks}. A \textit{MEHost} is denoted by $h \in \mathbb{H}$, and a \textit{HostLink} between two \textit{MEHosts} is denoted by $e \in \mathbb{E}$. The \textit{HostLinks} can be either physical or virtual links with fixed capacities and latencies.

Suppose in a certain service area, there are $n_s$ \textit{users} from various \textit{MECSPs} requesting the same chained service $s$ from a \textit{MEAV}. We use $m$ to denote a \textit{MECSP} and $h_m$ for a \textit{MEHost} that belongs to $m$. Define an assigning function $x_{v h_m}$, whose value is $1$ if VM $v$ is assigned to Host $h_m$, 0 otherwise. 
\begin{equation}
x_{v h_m} \triangleq \begin{cases}
1, & v \text{ is deployed on }h_m; \\ 
0, & \text{otherwise}.
\end{cases}
\end{equation}

Define a binary indicator of an \textit{AppLink} between two chained \textit{MEApps} in $s$, denoted by $L(v_{h_i}, v_{h_j})$, such that
\begin{equation}
\label{eqn:linkbin}
L(v, v') \triangleq \begin{cases}
1, & \text{$l\in \mathbb{L}_s$ exists between } v \text{ and } v'; \\ 
0, & \text{otherwise}.
\end{cases}
\end{equation}

Also, we use $e_{ij}$ to represent the \textit{HostLink} between $h_i$ and $h_j$. The cost of deploying $s$ is the sum of the cost of deploying each \textit{MEApp} $v$ of the service and the cost of the traffic between each two adjacent \textit{MEApps} in the service chain. It can be shown by
\begin{equation}
\begin{aligned}
c_s = & \sum_{h_m \in H} \sum_{v \in \mathbb{V}_s} c_{v h_m} x_{v h_m} \\
      & + \sum_{h_i, h_j \in H}\sum_{v, v' \in \mathbb{V}_s} c_{v h_{i}, v' h_{j}} x_{v h_i} x_{v' h_j} L(v, v'),
\end{aligned}
\end{equation}
where $c_s$ represents the cost of deploying $s$ and $c_{v h_m}$ is for the cost of a \text{MEApp} $v$ deployed on a \text{MEHost} $h_m$. We assume that the pricing scheme for the same \textit{MECSP} is the same across all of its hosts. For a \textit{MEHost} $h_m$, define its basic unit resource price, which is the unit price of serving its own subscribers, as $\gamma_m$. When $h_m$ is serving users of other \textit{MECSPs}, it charges a premium of $\delta_m$ for its unit resource, as the return for doing courtesy for its partners. Therefore, the shared unit resource price of $h_m$ can be represented by $(\gamma_m + \delta_m)$. Define $C_{h_m}$ and $M_{h_m}$ to be the capacity of vCPU and memory provided by $h_m$. Define $C_v$ and $M_v$ as the vCPU and memory consumed by $v$. Define $P_m$ to be the random variable for percentage of the \textit{users} using the service chain $s$ via networks of the \textit{MECSP} $m$. Depending on the numbers of active users for each \textit{MECSP}, the total cost for the \textit{MEAV} to place its \textit{MEApp} $v$ onto a host of $m$ is the cost incurred by \textit{users} of $m$ plus the cost by \textit{users} of other \textit{MECSPs}:
\begin{equation}
\begin{aligned}
c_{v h_m} = & n_s (C_v + M_v) P_m \gamma_m \\
            & + n_s (C_v + M_v) (1 - P_m) (\gamma_m + \delta_m) \\
          = & n_s (C_v + M_v) \left[ P_m \gamma_m  + (1 - P_m) (\gamma_m + \delta_m) \right] \\
          = & n_s (C_v + M_v) \left[ \gamma_m + (1 - P_m) \delta_m \right].
\end{aligned}
\end{equation}

When a request from a \textit{user} for a service chain arrives, the blockchain would know the \textit{MECSP} from which the \textit{user} subscribes. For the same placement decision, the value $c_s$ can significantly differ over changing distribution of \textit{users}. An example can be two \textit{MECSPs} $m_1$ and $m_2$, each with one host $h_{m_1}$ and $h_{m_2}$. If all \textit{users} are subscribers of $m_1$ and all \textit{MEApps} are placed on \textit{MEHosts} of $m_1$, then the cost payable by the \textit{MEAV}s would be lower than if all \textit{users} were subscribers of $m_2$.

\begin{figure}[htb]
\centering
\includegraphics[width=.48\textwidth]{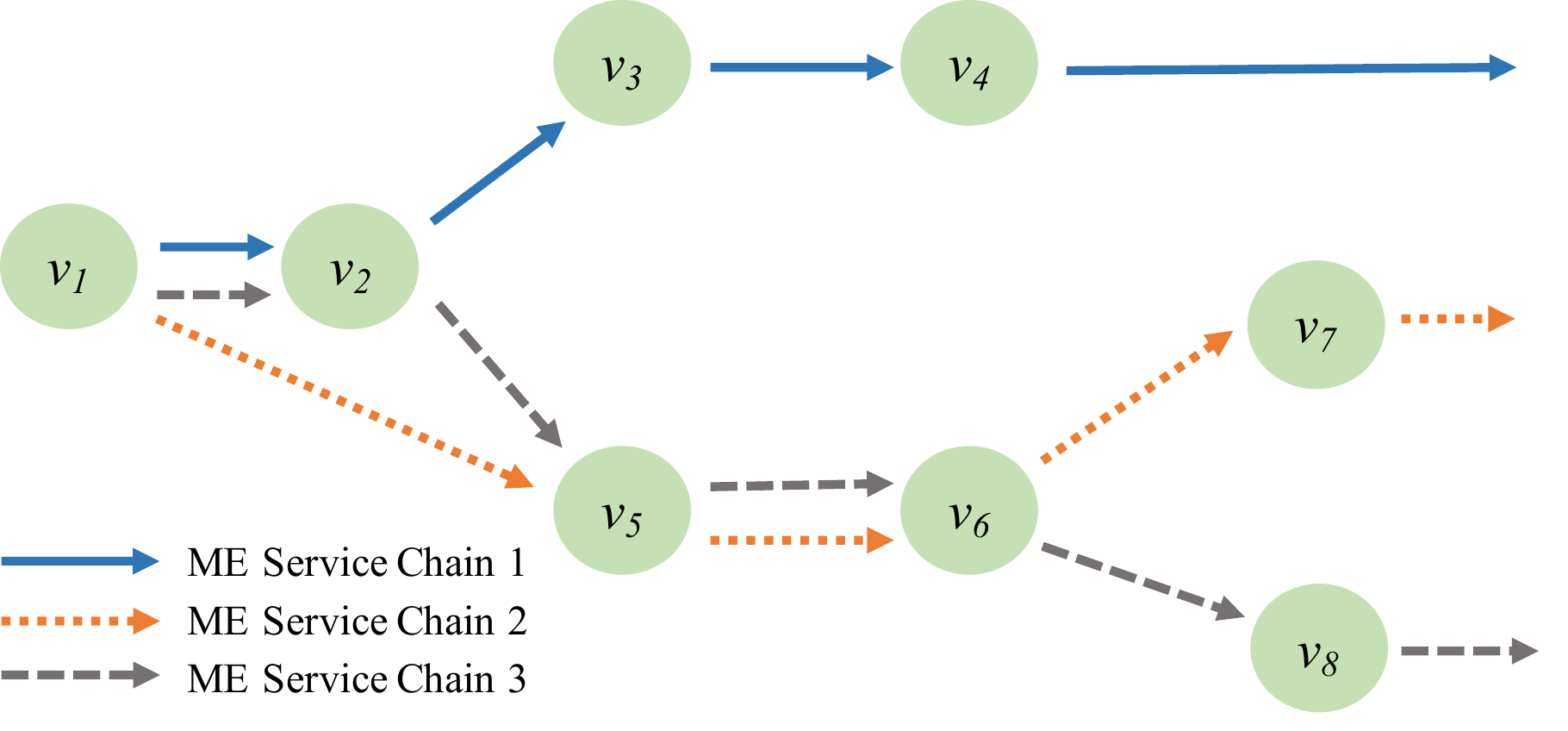}
\caption{\textbf{An example of ME service chains. There are 3 ME service chains sharing the services provided by 8 \textit{MEApps}.}}
\label{fig:sfc-example}
\end{figure}

\subsection{HostLink Unit Price}
When it comes to the cost modeling of a link between two \textit{MEHosts}, link availability is an important part for the service consistency of the \textit{MEApps}. If a heavily used \textit{HostLink} is down, consequences can be catastrophic: even if all individual \textit{MEApps} are running, the traffic would not be able to flow through between one or more pairs of \textit{MEApps} and the service chain would not be functional. For each \textit{HostLink} $e_{ij}$, there can be one or more \textit{AppLinks} sharing its bandwidth. \textit{HostLink} outages require migrating the \textit{MEApps} if they cannot be fixed in time. Therefore, \textit{HostLink} availability has significant influence on possible \textit{MEApp} migrations and the potential costs incurred.

The link unit price of a \textit{HostLink} $e_{ij}$, denoted by $\zeta_{e_{ij}}$, is then defined to describe how much to use the \text{HostLink} $e_{ij}$. The following two parameters will determine $\zeta_{e_{ij}}$.

The first parameter is $L(v_{h_i}, v_{h_j})$ as defined in Eqn. (\ref{eqn:linkbin}). The more \textit{AppLinks} a \textit{HostLink} carries, the more vital and expensive it becomes. The reason behind this ranking parameter is the potential consequence of migration: failure of a \textit{HostLink} used by many VMs would lead to massive migration of all \textit{MEApps} connected by that \textit{HostLink}, which would be more disruptive to the service chain.

The other parameter $B_V(e_{ij})$ is the total bandwidth consumed by traffic between \textit{MEApps} on the two hosts. It is selected because larger bandwidth usages would cause challenges at the time of migration: it can be hard to find another link with enough capacity.

\begin{equation}
B_V(e_{ij}) \triangleq \left[ \sum_{v_{h_i}, v_{h_j}, h_i \neq h_j} B(v_{h_i}, v_{h_j}) \right].
\end{equation}

Combining the two parameters, we define the unit price $\zeta_{e_{ij}}$ of a \textit{HostLink} $e_{ij}$, as the factor of the number of \textit{AppLinks} between two hosts times the factor of traffic flowing through these links:
\begin{equation}
\zeta_{e_{ij}} = \frac{\left[ \sum_{v_{h_i}, v_{h_j}, h_i \neq h_j} L(v_{h_i}, v_{h_j}) \right]}{N_{e_{ij}}} \frac{B_V(e_{ij})}{B(e_{ij})},
\end{equation}
where $N_{e_{ij}}$ is the maximum number of virtual links possible on $e_{ij}$. Therefore, $\zeta_{e_{ij}} \in [0, 1]$. The value of $\zeta_{e_{ij}}$ will rise to mark up a link's importance given it is either occupied by more pairs of VMs, or there is more traffic assigned to $e_{ij}$, or both. The cost of any two \textit{MEApps} is then the sum of the cost serving users that belong to the \textit{MECSPs} owning $h_i$ and $h_j$ and the cost serving other users timed by the price factor $\kappa_{e_{ij}}$:

\begin{equation}
\begin{aligned}
c_{v h_{i}, v' h_{j}} = & n_s \zeta_{e_{ij}} (P_{m_{h_i}} + P_{m_{h_j}})\kappa_{m_{h_i} m_{h_j}} \\
      & + n_s \zeta_{e_{ij}} (1 - P_{m_{h_i}} - P_{m_{h_j}}) (\kappa_{m_{h_i} m_{h_j}} + \sigma_{m_{h_i} m_{h_j}}) \\
    = & n_s \zeta_{e_{ij}} [ (P_{m_{h_i}} + P_{m_{h_j}}) \kappa_{m_{h_i} m_{h_j}} \\
      & + (1 - P_{m_{h_i}} - P_{m_{h_j}}) (\kappa_{m_{h_i} m_{h_j}} + \sigma_{m_{h_i} m_{h_j}}) ].
\end{aligned}
\end{equation}

\subsection{HostLink latency}
Define the latency of the link $e_{ij}$ to be $t_{e_{ij}}$. For a service chain $s$, the total latency $t_s$ is then
\begin{equation}
t_s = \sum_{h_i, h_j \in \mathbb{H}} \sum_{v_{h_i}, v_{h_j} \in \mathbb{V}_s} L(v_{h_i}, v_{h_j}) x_{v h_i} x_{v h_j} t_{e_{ij}}.
\end{equation}
In the equation above, $t_{e_{ij}}$ is a constant depending on the particular $e_{ij}$. If $h_i = h_j$, then we consider the latency to be 0, since no actual \textit{HostLink} is used for data transmission between the two \textit{MEApps}. Define the maximum latency allowed for the service chain $s$ is $T_s$. Then there must be $t_s \leq T_s$ to meet the latency requirement.

\subsection{Stochastic Programming Formulation}

The problem is formulated as a stochastic programming optimization. Define $\mathbb{V}_h$ as the set of all \textit{MEApps} deployed on the \textit{MEHost} $h$. The objective is to minimize the total cost of the service chain $s$ to provide service with the lowest cost to the end user. As discussed in Section \ref{sec:problem}, the optimization is to minimize the costs on \textit{MEHosts} and \textit{HostLinks} for all \textit{MEApps} of $s$.

\noindent{\textit{Minimize}}
\begin{equation}
\label{eqn:objective}
\begin{aligned}
c_s = &\sum_{h_m \in \mathbb{H}} \sum_{v \in \mathbb{V}_s} c_{v h_m} x_{v h_m} \\
      &+ \sum_{h_i, h_j \in \mathbb{H}}\sum_{v, v' \in \mathbb{V}_s} c_{v h_{i}, v' h_{j}} x_{v h_i} x_{v' h_j} L(v, v') \\
    = &\sum_{h_m \in \mathbb{H}} \sum_{v \in \mathbb{V}_s} x_{v h_m} n_s (C_v + M_v) \left[ \gamma_m + (1 - P_m) \delta_m \right] \\
      & + \sum_{h_i, h_j \in \mathbb{H}}\sum_{v, v' \in \mathbb{V}_s} n_s \zeta_{e_{ij}} [ (P_{m_{h_i}} + P_{m_{h_j}}) \kappa_{m_{h_i} m_{h_j}} \\
      & + (1 - P_{m_{h_i}} - P_{m_{h_j}}) (\kappa_{m_{h_i} m_{h_j}} + \sigma_{m_{h_i} m_{h_j}}) ] L(v, v'),
\end{aligned}
\end{equation}
\begin{equation}
\label{eqn:constraint1}
\begin{aligned}
& \mathit{w.r.t.} \qquad x_{v h_m}, \\
& \mathit{s.t.} \qquad B(e_{ij}) \geq \sum_{v_{h_i}, v_{h_j}, h_i \neq h_j} B(v_{h_i}, v_{h_j}),
\end{aligned}
\end{equation}

\begin{equation}
\label{eqn:constraint2}
C_h \geq \sum_{v \in \mathbb{V}_h} C_v,
\end{equation}

\begin{equation}
\label{eqn:constraint3}
M_h \geq \sum_{v \in \mathbb{V}_h} M_v,
\end{equation}

\begin{equation}
\label{eqn:constraint4}
\sum_{h_i, h_j \in \mathbb{H}} \sum_{v_{h_i}, v_{h_j} \in \mathbb{V}_s} L(v_{h_i}, v_{h_j}) x_{v h_i} x_{v h_j} t_{e_{ij}} \leq T_s.
\end{equation}

\subsection*{\textbf{Remarks}}

\begin{itemize}
  \item Function (\ref{eqn:objective}) is the objective function. It minimizes the cost of all \textit{MEApps} and \textit{AppLinks} by using less hosts, while not exhausting them.
  \item Constraint (\ref{eqn:constraint1}) is the \textit{HostLink} bandwidth capacity bounds between each two hosts. Traffic transmitted between any two hosts $h_i$ and $h_j$ must not exceed the corresponding bandwidth capacity $B(e_{ij})$.
  \item Constraints (\ref{eqn:constraint2}) and (\ref{eqn:constraint3}) are the CPU and memory capacity bounds for each \textit{MEHost}. The CPU and memory used by \textit{MEApps} coordinating with each other and by intra-host communications must not exceed $C_h$ and $M_h$.
  \item Constraint (\ref{eqn:constraint4}) is the latency requirement of the service chain $s$ to ensure that the total latency of $s$ must not exceed the maximum latency allowed $T_s$.
\end{itemize}

\section{The \textsf{EdgeChain} Placement Algorithm}
\label{sec:algorithm}
The formulation presented in the previous section is a stochastic programming problem. Problems of this type been proved to be NP-hard \cite{gaivoronski2011knapsack}. It may not be computationally feasible when attempting to solve it in large scale. To apply our model to real-world scenarios, we design a heuristic algorithm called \textsf{EdgeChain} to achieve suboptimal results by applying a hybrid strategy of best-fit and first-fit decreasing algorithm. The pseudo code of the algorithm is shown in Algorithm \ref{algo:edgechain}.

\begin{algorithm}[htb]
    \DontPrintSemicolon
    \SetKwData{HostList}{host\_list}
    \SetKwData{CurrMEApp}{app}
    \SetKwData{RemainingCPU}{cpu\_left}
    \SetKwData{RemainingMem}{mem\_left}
    \SetKwData{Latency}{latency}
    \SetKwData{MaxLatency}{max\_latency}
    \SetKwData{And}{{\bf and}}\SetKwData{Or}{{\bf or}}
    \KwData{\HostList: list of candidate MEHosts}
    \KwData{\CurrMEApp: requested \textit{MEApp} to be placed, including its max latency allowed, stored in \Latency}
    \KwData{\MaxLatency: max latency allowed for the service chain}
    \KwResult{The best \textit{MEHost} in \HostList to place \CurrMEApp, or \textsf{none} if no valid host is found}
    \Begin{
      sort by percentage of \textit{users} of the service chain descending \;
      \If{multiple \textit{MEHosts} found} {
        sort \HostList by the locations of \CurrMEApp's last-hop \textit{MEApps} \;
        \If{still multiple \textit{MEHosts} found} {
          sort by the latency of the \textit{HostLinks} to the previous \textit{MEApps} in the service chain ascending \;
        }
      }
      \For{$h \in \HostList$} {
        $\Latency \leftarrow $ all latencies added together if \CurrMEApp placed on $h$ \;
		\If{\Latency $\leq$ \MaxLatency} {
	      \RemainingCPU $\leftarrow$ calculate remaining vCPU by $C_h$ and $C_v$ of each \textit{MEApp} placed on $h$ \;
		  \RemainingMem $\leftarrow$ calculate remaining memory by $M_h$ and $M_v$ of each \textit{MEApp} placed on $h$ \;
		  \If{$\RemainingCPU \geq 0$ \And $\RemainingMem \geq 0$} {
		    \Return $h$\;
		  }
		}
      }
      \Return \textsf{none}\;
    }
    \caption{\textsf{EdgeChain} Placement Algorithm}
  \label{algo:edgechain}
\end{algorithm}

\subsection{Processing Order and selection of MEHosts}
The \textsf{EdgePlace} algorithm runs on each mining node based on the Ethereum platform. The algorithm retrieves its input information from the blockchain, as all transactions and updates are recorded on the blockchain. The \textsf{EdgePlace} algorithm will select the \textit{MEHosts} following the steps below.

\subsubsection{Users}
Sort all \textit{MEHosts} by the percentage of \textit{users} of the service chain. For each \textit{MEApp} on the service chain, consider which \textit{MECSP} has most users using it. Then \textit{MEHosts} with the same \textit{MECSP} will have higher ranks to deploy this \textit{MEApp}. Since all \textit{MEHosts} of the same \textit{MECSP} have the same unit resource cost, the \textit{MEApp} can be placed on any of the \textit{MEHosts} that belongs to the best \textit{MECSP}, to avoid the situation that too many \textit{MEApps} are concentrated on one \textit{MEHost}.

\subsubsection{Last-hop MEApp}
For \textit{MEHosts} given higher priority in the previous step, sort by the locations of last-hop \textit{MEApps}. \textit{MEHosts} hosting the previous-hop \textit{MEApps} will be considered first. This step is to reduce the traffic cost between different \textit{MECSPs}.

\subsubsection{Latency}
For \textit{MEHosts} given higher priority in the previous step, sort by the latency of the \textit{HostLinks} to the previous \textit{MEApps} in the service chain. \textit{MEHosts} with lower latency will be considered first.

After the list of candidate \textit{MEHosts} are sorted according to the steps above, the algorithm iterates the list and pick the first valid \textit{MEHost} that has enough resources to place the \textit{MEApp}, as well as meeting the latency requirement of the service chain.

\section{EdgeChain Design and Implementation}

In this section, we introduce the design and implementation of \textsf{EdgeChain}, a blockchain-based system that integrates with the existing MEC architecture for \textit{MECSPs} and the scheduler of \textit{MEAV}. There are mainly two reasons the blockchain is used in the system:
\begin{itemize}
  \item The blockchain acts as a public ledger that stores all useful information and transactions made during the placement process. Exposure of the information would help the placement algorithm make optimized decisions considering the global resource demand and allocation. The blockchain enables such centralized resource information, in a decentralized implementation.
  \item As a public ledger applying proof-of-work verifications, the blockchain makes it nearly impossible to tamper the history stored in the blockchain. The \textsf{EdgeChain} algorithm will be downloaded by all mining nodes and they will execute the same algorithm with the same input. The placement result will only be accepted by the system if majority of the mining nodes reach agreement on the output. This will ensure the neutrality of the placement decisions.
\end{itemize}

The system takes requests to place \textit{MEApps} from \textit{MEAVs}, and the placement algorithm runs as the smart contract on the blockchain to select the best \textit{MEHost} from all candidates. The NFV orchestrator of the related \textit{MECSP} receives and enforces the placement decision, while posting the transaction onto the blockchain for recording. While this paper is written, the blockchain is implemented based on VeChain \cite{vechain}, an enterprise-level blockchain-as-a-service framework derived from Ethereum.

\begin{figure}[htb]
\centering
\includegraphics[width=.48\textwidth]{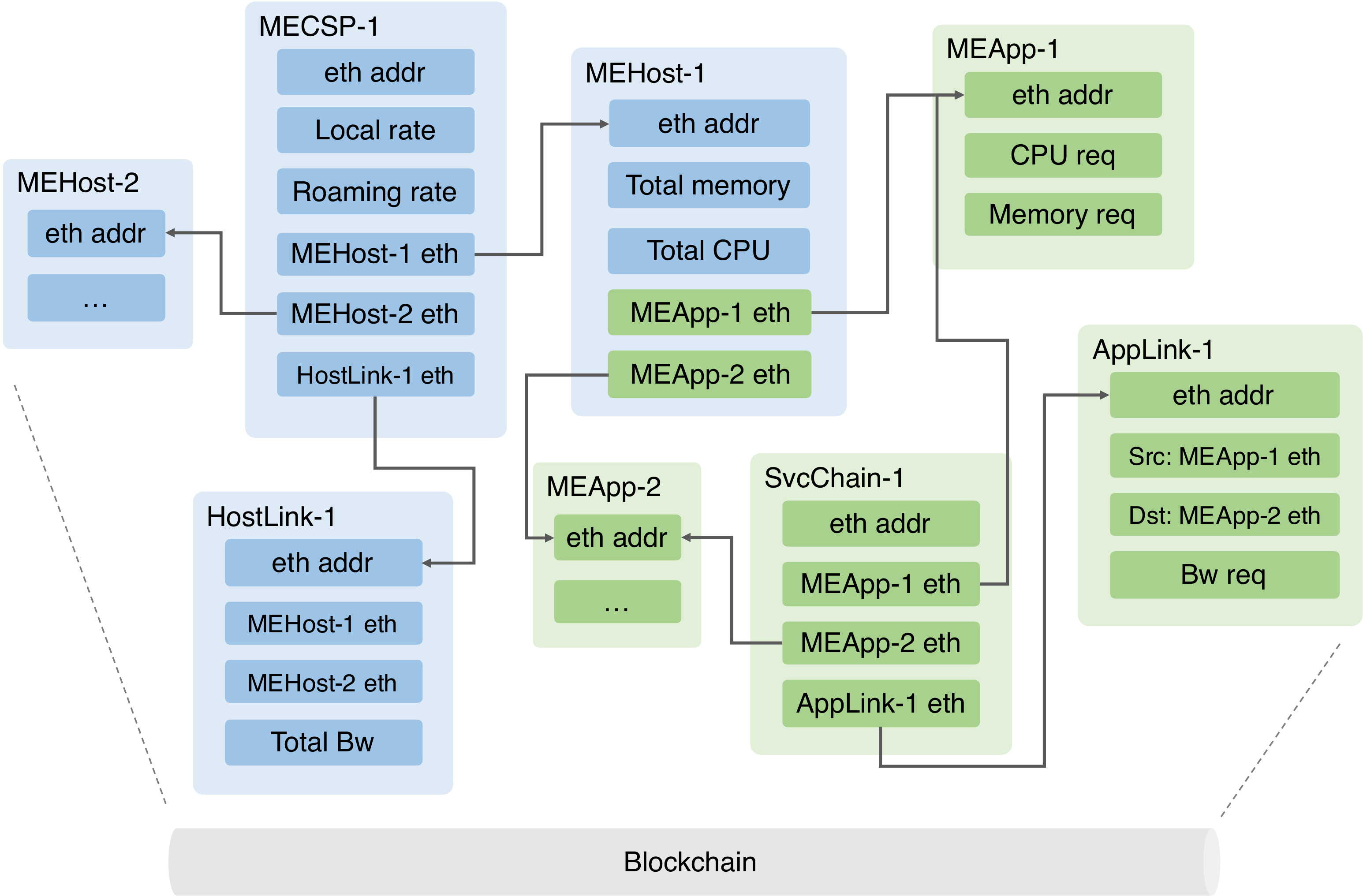}
\caption{\textbf{Data entities and their relationship used by and stored in EdgeChain, including \textit{MECSPs}, \textit{MEHosts}, \textit{HostLinks}, \textit{SvcChains}, \textit{MEApps}, and \textit{AppLinks}.}}
\label{fig:data-structure}
\end{figure}

\subsection{Data Entities}

As Fig. \ref{fig:data-structure} shows, there exist 6 types of data entities on the blockchain and they are related to each other to represent the status of \textit{MEHosts} and placement decision of running \textit{MEApps}. The descriptions of these data entities are illustrated below. Each data entity record has a unique Ethereum address for other to locate it on the blockchain. All types of data entities can be created, updated and deleted, while the blockchain will keep the audit trail of every change.

\subsubsection{MECSP}
When a \textit{MECSP} record is registered to \textsf{\textsf{EdgeChain}}, a record of this \textit{MECSP} is added with the Ethereum addresses pointing to the records of all its eligible \textit{MEHosts} and \textit{HostLinks}. A \textit{MECSP} record is updated whenever there is change to any \textit{MEHost} or \textit{HostLink}.

\subsubsection{MEHost}
A \textit{MEHost} record registers under an existing \textit{MECSP} to the blockchain. In a record, the vCPU and memory capabilities can be found, along with the Ethereum addresses pointing to the records of all \textit{MEApps} placed onto it.

\subsubsection{HostLink}
Similar to \textit{MEHosts}, a \textit{HostLink} is under a registered \textit{MECSP}, which contains the two \textit{MEHosts} it connects, and the bandwidth of the \textit{HostLink}.

\subsubsection{SvcChain}
A service chain is registered by a \textit{user} to the blockchain to reflect the resource consumption of a chained service, including that from \textit{MEApps} and the corresponding \textit{AppLinks}. The service chain can have \textit{MEApps} from multiple \textit{MEAVs}.

\subsubsection{MEApp}
A \textit{MEAV} will submit a record of a \textit{MEApp} whenever it needs to spin up one. A record stores the vCPU, memory usage of the \textit{MEApp}.

\subsubsection{AppLink}
\textit{AppLinks} describe chained relationship between two \textit{MEApps}. The source and destination \textit{MEApps} are stored in an \textit{AppLink} record, as well as network bandwidth requirement of this link.

\begin{figure}[htb]
\centering
\includegraphics[width=.48\textwidth]{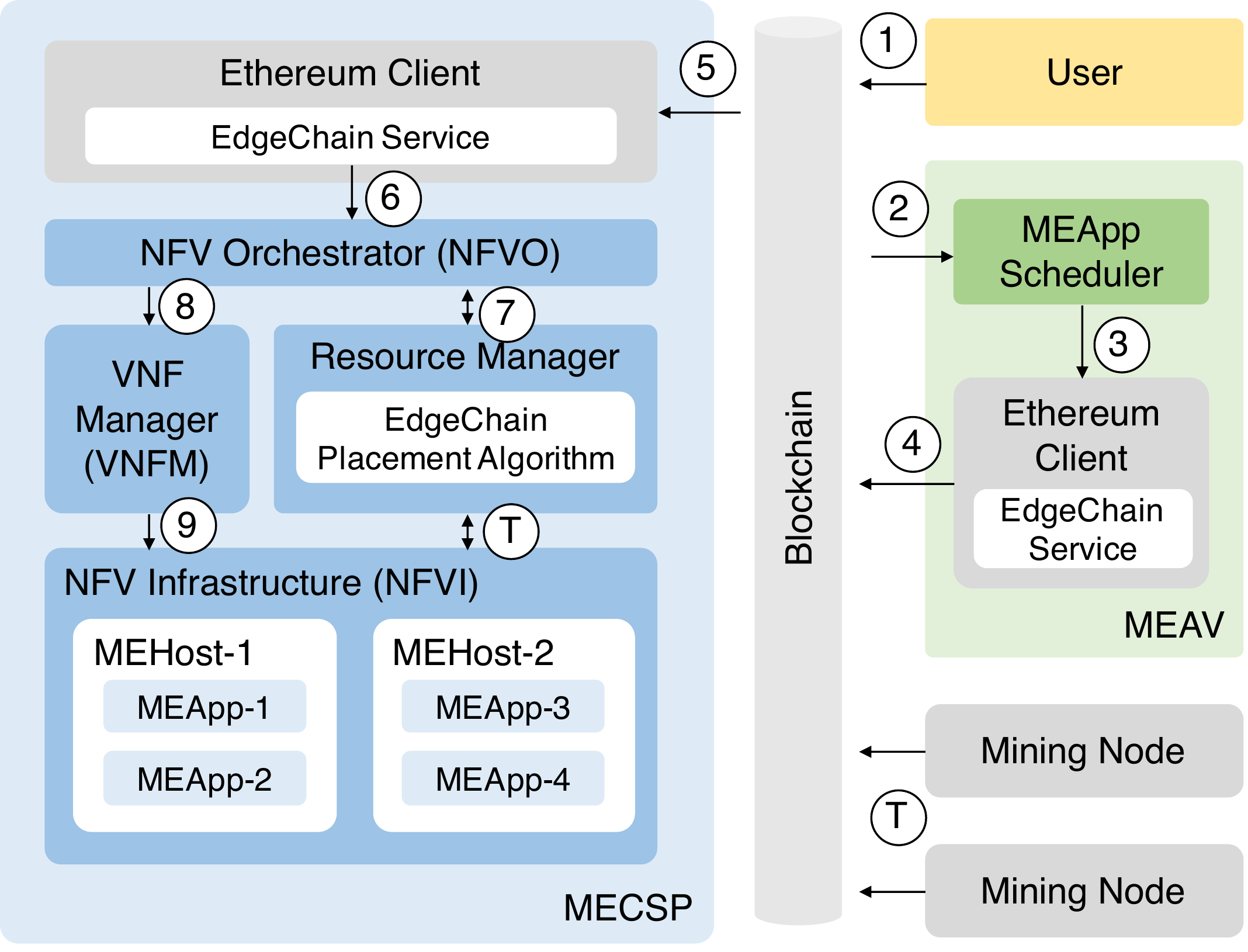}
\caption{\textbf{Typical work flow of EdgeChain. \textit{MECSPs}, \textit{MEAVs}, and mining nodes participate in the process. Steps of the work flow are marked by circled numbers and alphabets with details documented in Section \ref{sec:workflow}.}}
\label{fig:workflow}
\end{figure}

\subsection{EdgeChain Work Flow}
\label{sec:workflow}
A typical \textsf{EdgeChain} work flow can be demonstrated by Fig. \ref{fig:workflow}, where there are three parties participating in the entire process: \textit{MECSPs}, \textit{MEAVs}, and mining nodes. We use circled numbers and alphabets to define the work flow in sequence. 

\circled{1} A \textit{user} requests a service chain from the blockchain. Such requests will be sent to the blockchain every time a user requests a service chain.

\circled{2} The request for the service chain is recorded. When the request is synced to the mining nodes, it will be broken into requests for \textit{MEApps}. The mining nodes will run the logic to break down the service chain creation request. Then the requests for \textit{MEApps} are propagated to all corresponding \textit{MEAVs}.

\circled{3} Based on its user demand, the \textit{MEApp} Scheduler decides to create a new instance of \text{MEApp} and pass the request to the Ethereum client of the \textit{MEAV}.

\circled{4} The Ethereum client running the \textsf{EdgeChain} service sends the request to the blockchain, creating records for the request of placing a new \textit{MEApp}.

\circled{5} The request of creating a new \textit{MEApp} arrives at a MECSP through its Ethereum client. 

\circled{6} For every \textit{MECSP}, the Ethereum client requests the NFV Orchestrator (NFVO) to call the \textsf{EdgeChain} placement algorithm downloaded to the resource manager for the decision of the placement. This will ensure that the placement algorithm be executed by different parties for verifying the results. The placement result returned by the next step will only be accepted if majority of the parties return the same placement result.

\circled{7} The NFVO calls the \textsf{EdgeChain} placement algorithm for the placement decision. Note that the decision can be a hash representing any \textit{MEHost} within the entire MEC network. If the result points to a \textit{MEHost} which does not belong to the current \textit{MECSP}, then no actual placement will be done. Instead, only the result along with the algorithm's hash will be returned to the Ethereum client for verification.

\begin{figure}[htb]
\centering
\includegraphics[width=.48\textwidth]{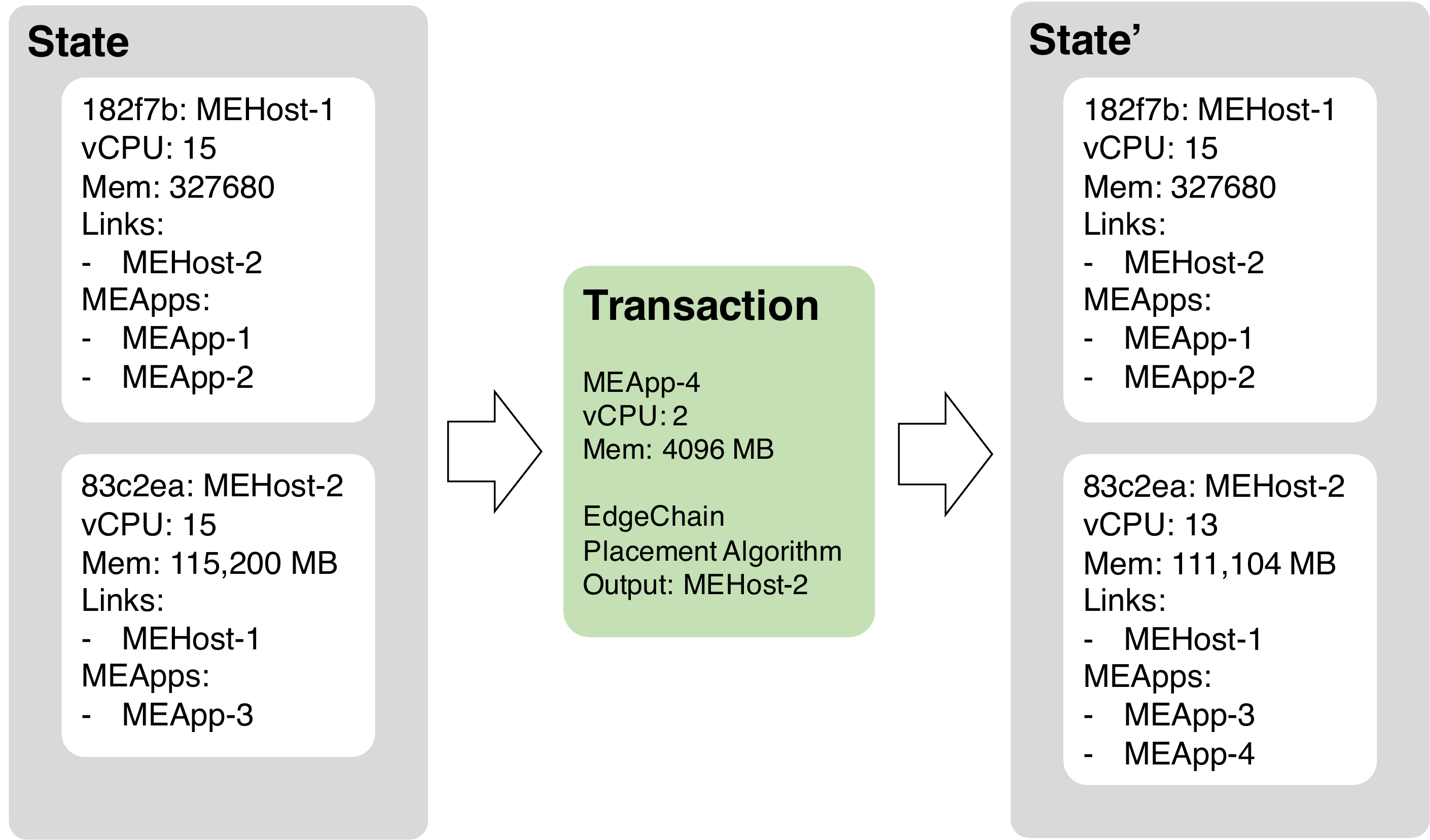}
\caption{\textbf{A placement transaction in EdgeChain. A state transition happens upon a transaction. As this figure shows, \textit{MEApp-4} is to be placed with the requirement of 2 vCPUs and 4096 MB of memory. The input of the EdgeChain placement algorithm is the current state of the two \textit{MEHosts}. The result is to place \textit{MEApp-4} onto \textit{MEHost-2}. After the transaction is accepted, the resources taken by \textit{MEApp-4} are deducted from the remaining resources of \textit{MEHost-2}.}}
\label{fig:state}
\end{figure}

\circled{8} If the result points to a \textit{MEHost} of the current \textit{MECSP}, then the NFVO will sends the request to place the \textit{MEApp} to the VNF Manager (VNFM). Also, a transaction shown in Fig. \ref{fig:state} will be posted to the blockchain to record that placement actually occurs.

\circled{9} The VNFM sends the request to the NFV Infrastructure (NFVI) deploy the \textit{MEAPP} onto the target \textit{MEHost}.

\circled{T} The mining nodes periodically perform the mining process to verify the blockchain, as well as earning Ethers for requesting placement services. Meanwhile, the resource manager periodically synchronizes with the NFVI for the up-to-date resource usage and availability, and then posts the updated information to the blockchain.

\section {Numerical Results}
\label{sec:numerical}

In this section, we illustrate the numerical results of the MEC placement cost changes based on varying mobile edge application user cases using CloudSim \cite{cloudsim}. To clearly demonstrate the focused trends, the following assumptions are made to simplify the modeling of the problem without losing generality. We first discuss the placement results output by the \textsf{EdgeChain} algorithm for the same service chain on the same set of \textit{MEHosts}.
\begin{enumerate}
  \item The unit costs of the CPU and memory of all hosts for the same \textit{MECSP} are the same.
  \item Costs of network bandwidth for all links follow the same unit price.
  \item One mobile edge application includes the same type of VMs with the same CPU, memory and network bandwidth requirements.
  \item A request from the user will be processed by one VM, while the VM may communicate with other VMs to exchange information.
\end{enumerate}

\subsection{Parameters}
With the assumptions above, we choose parameters for our placement model to evaluate the performance and the facts under different circumstances. First, we choose a MEC service scenario of 3 \textit{MECSPs} $m_1$, $m_2$, and $m_3$, each with 3 \textit{MEHosts}, where $h_1$, $h_2$, $h_3$ belong to $m_1$, $h_4$, $h_5$, $h_6$ belong to $m_2$, and $h_7$, $h_8$, $h_9$ belong to $m_3$. 

Three identical requested service chain, each with 5 \textit{MEApps} is to be placed. The \textit{MEApps} of each service chain are denoted by $v_1$, $v_2$, $v_3$, $v_4$, and $v_5$. The service chain starts from $v_1$ and ends at $v_5$: $v_1 \rightarrow v_2 \rightarrow v_3 \rightarrow v_4 \rightarrow v_5$. We assume that all \textit{MEApps} have the same CPU, memory and bandwidth requirements, which are shown in Table \ref{table:params}, along with other parameters. 

\begin{table}[htb]
  \caption{Parameters for the MEC scenario}
  \begin{tabularx}{.48\textwidth}{XX|XX}
  \toprule %
  \textbf{Parameter} & \textbf{Value} & \textbf{Parameter} & \textbf{Value} \\
  \toprule %
  $C_v$            & $2$ vCPUs    & $M_v$             & $2048$ MB     \\
  $C_h$            & $64$ vCPUs   & $M_h$             & $65536$ MB    \\
  $\gamma_{m_1}$   & 1.0          & $\delta_{m_1}$    & 0.2          \\
  $\kappa_{m_1}$   & 1.0          & $\sigma_{m_1}$    & 0.2          \\
  $\gamma_{m_2}$   & 0.8          & $\delta_{m_2}$    & 0.5          \\
  $\kappa_{m_2}$   & 0.8          & $\sigma_{m_2}$    & 0.5          \\
  $\gamma_{m_3}$   & 1.2          & $\delta_{m_3}$    & 0.3          \\
  $\kappa_{m_3}$   & 1.2          & $\sigma_{m_3}$    & 0.3          \\
  $n_s$            & 100 users    & $P_m$             & \textit{var}  \\
  $B(e_{ij})$      & 10000 Mbps   & $B(v, v')$        & 30 Mbps       \\
  $t_{e_{ij}}$     & 15 ms        & $T_s$             & 50 ms         \\
  \bottomrule
  \end{tabularx}
  \label{table:params}
\end{table}

\begin{figure}[t!]
\centering
    \begin{subfigure}
    \centering
    \includegraphics[width=2.5in, trim={1in 2.58in 1in 2.58in}]{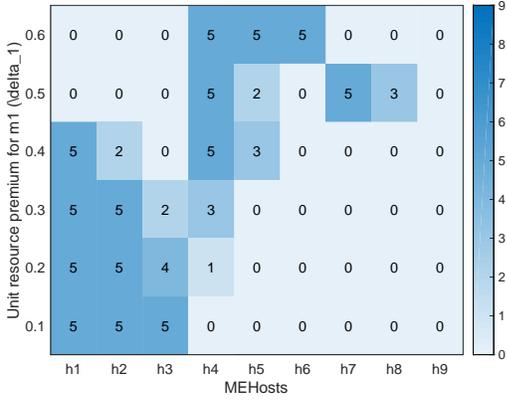}
    {\\(a) $P_{m_1} = 0.5$ and $P_{m_2} = P_{m_3} = 0.25$.}
    \end{subfigure}
    ~
    \begin{subfigure}
    \centering
    \includegraphics[width=2.5in, trim={1in 2.58in 1in 2.58in}]{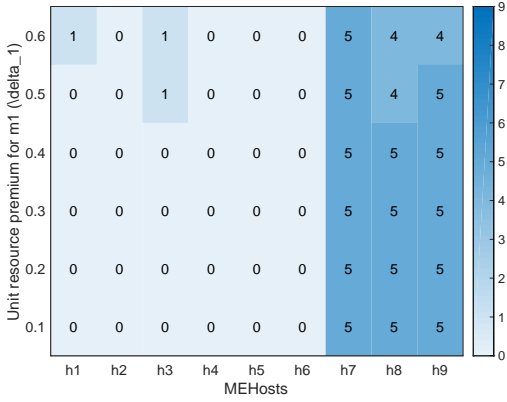}
    {\\(b) $P_{m_1} = P_{m_2} =0.25$ and $P_{m_3} = 0.5$.}
    \end{subfigure}
\caption{\textbf{Placement results of 3 service chains consisting of 15 \textit{MEApps} in all. The value of $\delta_{m_1}$ changes from 0.1 to 0.6. Figure (a) shows the placement decision when $P_{m_1} = 0.5$ and $P_{m_2} = P_{m_3} = 0.25$. In comparison, Figure (b) shows the placement decision when $P_{m_1} = P_{m_1} = 0.25$ and $P_{m_3} = 0.5$.}}
\label{fig:place}
\end{figure}

\subsection{Placement trends with changing unit resource premium}

The placement decision changes by the increase of $\delta_{m_1}$ under different user distributions are shown in Fig. \ref{fig:place}, where $\delta_{m_1}$, the unit resource premium payable to the \textit{MECSP} for hosting \textit{MEApps} for others, increases from 0.1 to 0.6. For comparison, in Fig. \ref{fig:place}(a), most users are from $m_1$. There is $P_{m_1} = 0.5$ and $P_{m_2} = P_{m_3} = 0.25$. Meanwhile, in Fig. \ref{fig:place}(b), most users subscribe services from $m_3$ as $P_{m_1} = P_{m_1} = 0.25$ and $P_{m_3} = 0.5$.

From the results of the two scenarios, we learn that the \textit{MEHosts} with lower combination of unit resource base price ($\gamma_m$) and unit resource premiums ($\delta_m$) will be selected first. The \textit{MEHosts} of the \textit{MECSP} will have more weight upon consideration if there are more \textit{users} from that \textit{MECSP}.

\begin{figure}[tb]
\centering
\includegraphics[width=.48\textwidth, trim={0.5in 4in 0.5in 4in}]{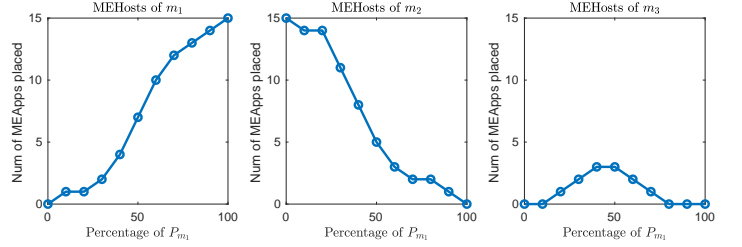}
\caption{\textbf{Numbers of \textit{MEApps} placed on the 3 \textit{MEHost} with different percentages of \textit{users} in the network. Users of $m_1$ increase from 0\% to 100\%, while those of $m_2$ decrease from 100\% to 0\%. There is no user for $m_3$.}}
\label{fig:user-dist}
\end{figure}

\subsection{Placement trends with changing user distribution}

To further demonstrate the impact from the distribution of the \textit{users}, we simulate various scenarios with different percentages of \textit{users} for $m_1$ and $m_2$, while there is no user for $m_3$. Users of $m_1$ increase from 0\% to 100\%, while those of $m_2$ decrease from 100\% to 0\%.

The results have shown the trends of \textit{MEApps} migrating to \textit{MEHosts} owned by the \textit{MECSP} that has more active users to avoid premiums charged by other \textit{MECSPs}. However, resource sharing still takes place ($m_3$ hosting \textit{MEApps} for $m_1$ and $m_2$) when needed for better latency results and service quality.

\section{Conclusions}
\label{sec:conclusion}
In this paper, we have presented the architecture and the algorithms for mobile edge applications placement for multiple mobile edge computing service providers, leveraging the blockchain-based system called \textsf{EdgeChain}. Future work will be considering multiple service chains initiated by multiple users, to achieve lower overall costs for the entire system.

\bibliographystyle{IEEEtran}
\bibliography{refs}

\end{document}